\documentclass[aps,prl,showpacs,superscriptaddress,notitlepage,twocolumn]{revtex4-1}

\usepackage{mathrsfs}
\usepackage{amsfonts}
\usepackage{amssymb}
\usepackage{amsmath}
\usepackage{bm}
\usepackage{graphicx}
\usepackage{sidecap}
\usepackage{color}
\usepackage[colorlinks=true,citecolor=blue,linkcolor=magenta]{hyperref}
\usepackage{subeqnarray}
\usepackage{verbatim}
\usepackage{ulem}
\usepackage{cases}
\usepackage{cancel}

\newcommand{\ket}[1]{\vert #1 \rangle}
\newcommand{\bra}[1]{\langle #1 \vert}

\begin{document}
\title{Topological Dynamical Decoupling}
\author{Jiang Zhang}
\affiliation{State Key Laboratory of Low-Dimensional Quantum Physics and Department of Physics, Tsinghua University, Beijing 100084, China}
\affiliation{Beijing Academy of Quantum Information Sciences, Beijing 100193, China}

\author{Xiao-Dong Yu}
\affiliation{Naturwissenschaftlich-Technische Fakult\"at, Universit\"at Siegen,
Walter-Flex-Stra\ss e 3, 57068 Siegen, Germany}

\author{Gui-Lu Long}
\email{gllong@tsinghua.edu.cn}
\affiliation{State Key Laboratory of Low-Dimensional Quantum Physics and Department of Physics, Tsinghua University, Beijing 100084, China}
\affiliation{Beijing National Research Center for Information Science and Technology, Beijing 100084, China}
\affiliation{Collaborative Innovation Center of Quantum Matter, Beijing 100084, China}
\affiliation{Beijing Academy of Quantum Information Sciences, Beijing 100193, China}

\author{Qi-Kun Xue}
\email{qkxue@tsinghua.edu.cn}
\affiliation{State Key Laboratory of Low-Dimensional Quantum Physics and Department of Physics, Tsinghua University, Beijing 100084, China}
\affiliation{Collaborative Innovation Center of Quantum Matter, Beijing 100084, China}
\affiliation{Beijing Academy of Quantum Information Sciences, Beijing 100193, China}
\begin{abstract}
We show that topological equivalence classes of circles in a two-dimensional square lattice can be used to design dynamical decoupling procedures to protect qubits attached on the edges of the lattice.
Based on the circles of the topologically trivial class in the original and the dual lattices, we devise a procedure which removes all kinds of local Hamiltonians from the dynamics of the qubits while keeping information stored in the homological degrees of freedom unchanged. If only the linearly independent interaction and nearest-neighbor two-qubit interactions are concerned, a much simpler procedure which involves the four equivalence classes of circles can be designed.
This procedure is compatible with Eulerian and concatenated dynamical decouplings, which make it possible to implement the procedure with bounded-strength controls and for a long time period.
As an application, it is shown that our method can be directly generalized to finite square lattices to suppress uncorrectable errors in surface codes.
\end{abstract}

\pacs{03.67.Lx, 03.67.Pp, 03.65.Vf.}

\keywords{Dynamical decoupling, Surface codes, Topological quantum computation}

\maketitle

\section{Introduction}
Protection of qubits from errors is a central and challenging task for quantum information processing \cite{di,ekert}.
One prominent approach for this aim is quantum error correction (QEC) which encodes logical states in a set of physical qubits to detect and correct errors, provided that the error rate of each operation on the qubits is below some threshold \cite{shor,steane96,laflamme96,terhal,yao12,barends14,nigg14,cor15,kelly15}.
A milestone of QEC is the invention of topological QEC, such as surface codes \cite{kitaev97,bravyi98,dennis02,fowler12pra,vijay,landau,brown,tuckett} and color codes \cite{bombin,kat,bombin15,lit,li}, in which quantum information is stored in topological degrees of freedom.
Unlike other QEC proposals, the qubits in topological QEC are placed on a particular lattice embedded in a surface.
For example, in surface codes \cite{kitaev97}, the qubits are arranged on a square lattice in a surface with holes, and the number of holes (or genus) determines how many logical qubits can be encoded \cite{bravyi98}.
A significant merit of introducing topological ideas to QEC is that it provides a modest requirement for error threshold, only about $7.4\times10^{-4}$ or even higher, for a single operation in surface codes \cite{fowler12prl,stephens14}.

Another route to protect qubits is to isolate them from the environment via dynamical decoupling (DD) \cite{viola98,viola99,viola00,viola03,viola05,kh05,kh10,uh07,gordon,yang,uys,west102,du09,zhang15,wang16}.
A pioneering work showed how to suppress dephasing on a single qubit by successively applying Pauli operations on it \cite{viola98}.
The idea is then extended to a general framework based on which linearly independent interactions between the qubits and their environment can be decoupled \cite{viola99, viola00}.
To date, there have been many DD proposals, including Eulerian DD \cite{viola03}, random DD \cite{viola05}, concatenated DD \cite{kh05,kh10}, and optimized sequences \cite{uh07,gordon,yang,uys,west102}.
However, none of the existing works on DD has considered where the qubits are placed and what we can benefit from the arrangement of the qubits.

In this work, we consider the case where a set of qubits interacts with their environment. Each of the qubits is attached to an edge of a two-dimensional square lattice embedded in a torus.
All the circles formed by the edges can be separated into four topological equivalence classes \cite{lidarbook}.
By using the circles belonging to the topologically trivial equivalence class in the original and dual lattices, we develop a DD procedure to remove all the unwanted interactions with the environment from the dynamics of the qubits.
When only the linearly independent interaction between the qubits and their environment and nearest-neighbor two-qubit interactions between the qubits are relevant, only two 4-ordered decoupling groups are needed in an alternative scheme,
where each element in these two groups is related to a different topological equivalence class in the original or dual lattice.
We explicitly show that this decoupling procedure can be realized with bounded-strength control Hamiltonians, and then generalize our method to planar square lattices which are used in a practical implementation of surface codes \cite{fowler12pra}.

\section{Square chains and their boundaries}
\begin{figure}[!tb]
  \centering
  \includegraphics[width=0.48\textwidth]{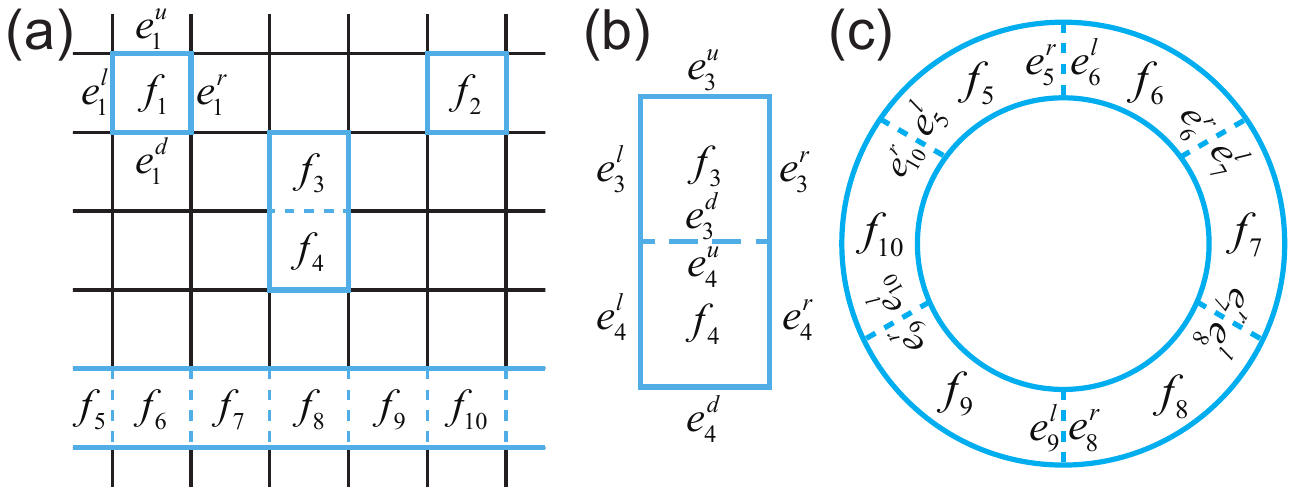}
  \caption{(a) Illustration of elements of $\mathcal{B}^1$ in a periodic $6\times6$ square lattice. When a chain has only one square, e.g., $f_1$, its boundary is the four surrounding edges (labeled by blue solid lines). (b) When a chain contains $f_3$ and $f_4$, the common edge (labeled by blue dashed line) shared by them is not included in the boundary since $e_3^d\!+\!e_4^u=0$. (c) When a set of squares ($f_5$ to $f_{10}$) forms a row in the lattice, they constitute a ring on the surface.}\label{fig1}
\end{figure}

Consider a square lattice attached on a torus (i.e., a periodic lattice) with $n$ rows and $n$ columns. Its faces (squares), edges, and vertices are labeled with $f$, $e$, and $v$, respectively.
Each square has four edges surrounding it as its boundary.
A square chain comprises one or more squares which can be shown in the form of $c^2=\sum_if_i$,
where $i$ is the index for the faces involved in the chain.
The boundary of a square chain is the sum of the boundaries for all the involved squares, taking the addition rule $e_i^j+e_i^j=0$ into account. Several concrete examples are illustrated in Fig.~\ref{fig1}. The boundaries of all the square chains form an Abelian group $\mathcal{B}^1$ with empty set as its identity element and addition of the edges as its group law.

Since there are a total of $n^2$ squares in the lattice, the number of all square chains is $\sum_{i=0}^{n^2}C(n^2,i)=2^{n^2}$,
where $C(n^2,i)$ is the number of $i$-combinations from the given $n^2$ squares.
It is worthwhile to note that the boundary of the square chain which consists of all the $n^2$ squares is empty.
Explicitly, we have the relation $\partial^2(\sum_{i=1}^{n^2}f_i)=0$, where $\partial^2$ is the boundary operator which projects a square chain to its boundary.
This relation implies that the order of $\mathcal{B}^1$ is half of $\sum_{i=0}^{n^2}C(n^2,i)$, which is $2^{n^2-1}$.

It is obvious that each element of $\mathcal{B}^1$ is a circle in the lattice. In fact, all the circles in the lattice can be divided into four kinds of topological equivalence classes: the topologically trivial circles that are boundaries of some faces; the circles that surround the ``handle'' of the torus; the ones that encircle the genus of the torus; the ones that surround both the ``handle'' and the genus.
Apparently, the elements of $\mathcal{B}^1$ belong to the topologically trivial equivalence classes.

\section{Topological dynamical decoupling}
The physical model that we consider is a set of qubits interacting with their environment. The corresponding total Hamiltonian of the full quantum system reads $H_t=H_S+H_E+H_{S\!E}$, where $H_S$ is the qubits' Hamiltonian, $H_E$ is the environment Hamiltonian, and $H_{S\!E}=\sum_i\sigma_{\alpha}^i\otimes E^i$ [$\sigma_\alpha^i$ and $E^i$ are the Pauli-$\alpha$ ($\alpha=x,y$, or $z$) operator acting on the $i$th qubit and its corresponding environment operator, respectively] is the interaction between them.
Dynamical decoupling can remove $H_{S\!E}$ from the dynamics of the qubits with fast and strong pulses \cite{viola98,viola99}.
A typical decoupling procedure is composed of repetitive segments, each of which can be described with a notation $[g_m^\dag, \tau, g_mg_{m-1}^\dag,\cdots, g_3g_2^\dag,\tau,g_2g_1^\dag,\tau,g_1]$, where $g_i$ are unitary operators generated by the control pulses.
From right to left, this notation means that, apply $g_1$ to the qubits, let the system evolve freely for time $\tau$, then apply $g_2 g_1^\dag$ followed by another free evolution for time $\tau$ and so on.
A decoupling group is formed when $g_i$ form a finite group $\mathcal{G}=\{g_1,g_2,\cdots,g_m\}$.
Let the time scale associated with one such segment be $T_c$.
In the ideal limit of an arbitrarily short $T_c$, the dynamics of the qubits is transformed through a dynamically averaged operator of the form
\begin{equation}\label{effha}
  \prod_{\mathcal{G}}(A)=\frac{1}{|\mathcal{G}|}\sum_{i}g_i^\dag Ag_i,
\end{equation}
where $|\mathcal{G}|$ is the order of group $\mathcal{G}$, and $A$ is an arbitrary operator acting on $S$ \cite{viola99}.
It is clear that if a proper group is chosen so that $\prod_{\mathcal{G}}(\sigma_\alpha^i)=0$ for all $i$, the errors caused by $H_{S\!E}$ can be eliminated.

\begin{figure}[!tb]
  \centering
  \includegraphics[width=0.45\textwidth]{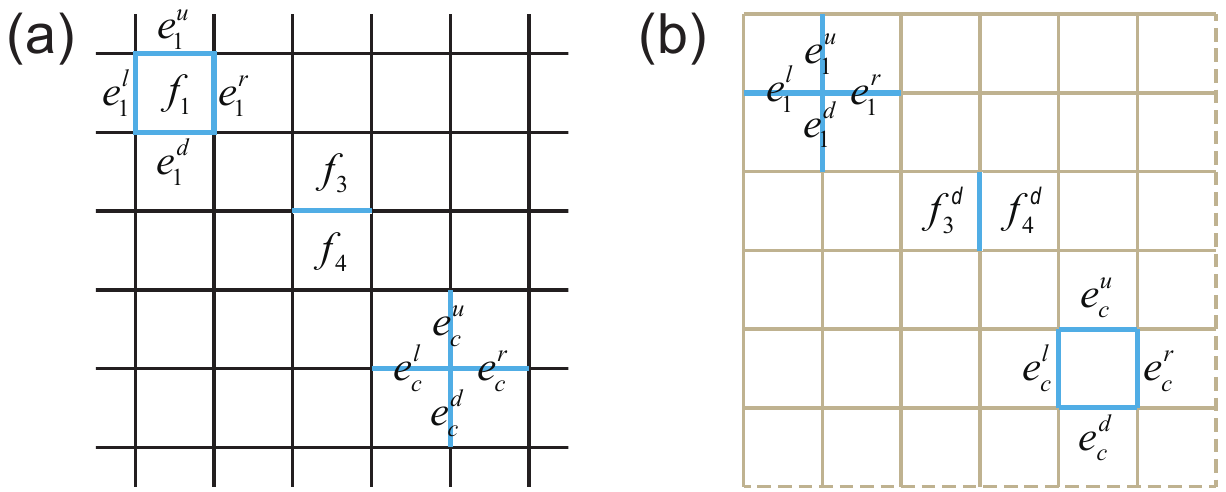}
  \caption{A periodic square lattice (a) and its dual (b). (a) The four edges ($e_1^u$, $e_1^l$, $e_1^d$, and $e_1^r$) connecting head-to-tail form the boundary of $f_1$. The four edges ($e_c^u$, $e_c^l$, $e_c^d$, and $e_c^r$) sharing the same vertex constitute a cross. (b) The square $f_1$ in the original lattice is transformed into a cross in the dual. Meanwhile, a cross in the original lattice is changed into a square in the dual.}\label{fig2}
\end{figure}

Now we show how to remove $H_{S\!E}$ from the dynamics of the qubits. We first define $\mathcal{B}^z$.
The elements of $\mathcal{B}^1$ and those of $\mathcal{B}^z$ are in a one-to-one correspondence.
An element of $\mathcal{B}^1$ contains some edges in the lattice and every edge supports a qubit.
An element $b^1_i\in \mathcal{B}^1$ can be transformed into its related element $b^z_i\in \mathcal{B}^z$ by replacing each edge of $b^1_i$ with the $\sigma_z$ operator acting on the corresponding qubit.
In this way, $b^z_i$ turn out to be unitary operators which are strings of Pauli-$Z$ operators on different qubits.

Group $\mathcal{B}^x$ is defined on the dual lattice shown in Fig.~\ref{fig2}(b).
The dual lattice is constructed by placing a vertex within each face, and connecting pairs of these vertices with an edge wherever the two corresponding faces have
overlapping boundaries. Each vertex of the original (or primal) lattice then corresponds to the face in the dual lattice whose boundary edges surround it.
As we did for the original lattice, we can construct a group $\mathcal{B}^{1d}$ whose elements are the boundaries of the square chains in the dual lattice.
Based on $\mathcal{B}^{1d}$, we can define a group $\mathcal{B}^{x}$ whose elements are in a one-to-one correspondence with  $\mathcal{B}^{1d}$'s elements.
An element $b^{x}_i\in \mathcal{B}^{x}$ can be obtained from its related element $b^{1d}_i\in \mathcal{B}^{1d}$ by replacing each edge of $b^{1d}_i$ with the $\sigma_x$ operator acting on the related qubit.

By using $\mathcal{B}^{xz}=\mathcal{B}^x\times\mathcal{B}^z$ as the decoupling group, we can eliminate local Hamiltonians (such as all $\sigma_\alpha^i$) from the dynamics with the decoupling procedure denoted by $[b_{2^{2n^2-2}}^{xz}, \tau, b_{2^{2n^2-2}}^{xz}b_{2^{2n^2-2}-1}^{xz},\cdots, b^{xz}_3b_2^{xz},\tau,b^{xz}_2b^{xz}_1,\tau,b^{xz}_1]$, where $b^{xz}_i$ are the elements in group $\mathcal{B}^{xz}$.
The reason is that group $\mathcal{B}^{xz}$ is an Abelian group, thus it has a total of $2^{2n^2-2}$ irreducible representations (irreps), each of which is 1-dimensional.
The corresponding group algebra takes the form of
\begin{equation}\label{}
  \mathbb{C}\mathcal{B}^{xz}\cong \oplus_{J=1}^{2^{2n^2-2}}M_{d_J},
\end{equation}
where $J$ is the irrep index, $d_J=1$ is the dimension of each irrep.
Due to the symmetry of the generators, the Hilbert space $\mathcal{H}$ spanned by the qubits can be decomposed into
\begin{equation}\label{}
  \mathcal{H}\cong\oplus_{J=1}^{2^{2n^2-2}} \mathbb{C}_{n_J}\otimes \mathbb{C}_{d_J},
\end{equation}
where all $n_J=4$ and $d_J=1$. The $n_J=4$ corresponds to the homology degrees of freedom.
Therefore, the DD procedure denoted by $[b_{2^{2n^2-2}}^{xz}, \tau, b_{2^{2n^2-2}}^{xz}b_{2^{2n^2-2}-1}^{xz},\cdots, b^{xz}_3b_2^{xz},\tau,b^{xz}_2b^{xz}_1,\tau,b^{xz}_1]$ can filter out all the Hamiltonians that have no component in the $\mathbb{C}_{n_J}$ section according to the dynamically averaged operator $\prod_{\mathcal{B}^{xz}}(A)$.
Apparently, a local Hamiltonian $H$ acting on the qubits, e.g., the single particle Hamiltonians and two-qubit interactions, has no component in $\mathbb{C}_{n_J}$, and thus $\prod_{\mathcal{B}^{xz}}(H)=0$.
As a result, the linearly independent interaction $H_I=\sum_{i,\alpha}\sigma_\alpha^i\otimes E_{\alpha}^i$ ($i$ is the qubit index and $\alpha=x,y,z$) between the qubits and the environment can be removed from the dynamics of the qubits.

\begin{figure}[!tb]
  \centering
  \includegraphics[width=.45\textwidth]{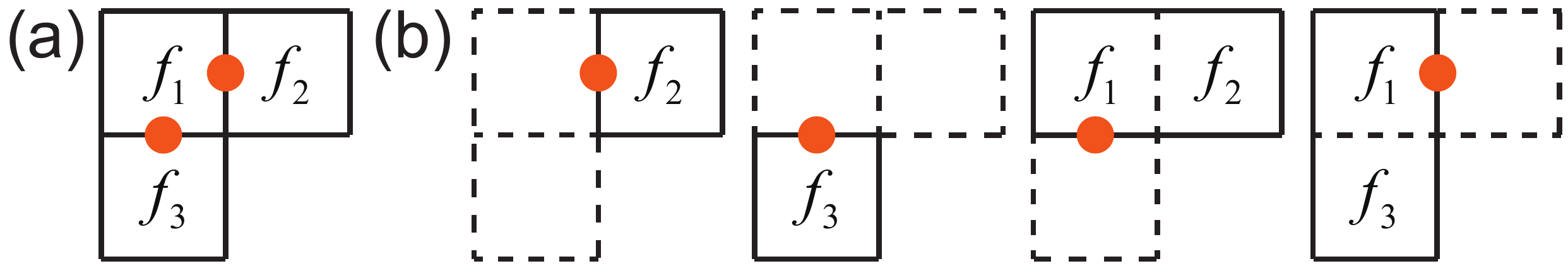}
  \caption{(a) A pair of nearest-neighbor qubits $a$ and $b$ (labeled as red circles), and related squares. Qubit $a$ is on the common edge of $f_1$ and $f_2$ while qubit $b$ is on the common edge of $f_1$ and $f_3$. (b) Four cases where the corresponding elements in $\mathcal{B}^z$ are not commutative with $H_{ab}$.}\label{fig3}
\end{figure}
In addition to the interaction $H_I$, our topological DD scheme can also remove the nearest-neighbour interactions between qubits. This is quite relevant to quantum error correction because the unwanted interactions between qubits can spread errors from one qubit to another. Since the error spread can cause undetectable errors, the elimination of the unwanted interactions can help to keep the errors local so that the undetectable errors are avoided.

Now we consider a pair of nearest-neighbor qubits $a$ and $b$ with a Heisenberg interaction $H_{ab}=\sum_{i=x,y,z}\sigma_i^a\sigma_i^b$ between them.
Among all the elements of $\mathcal{B}^z$, there are four kinds of elements that are not commutative with $H_{ab}$: those associated with $f_2$ but not $f_1$ and $f_3$; those associated with $f_3$ but not $f_1$ and $f_2$; those associated with $f_1$ and $f_2$ but not $f_3$; those associated with $f_1$ and $f_3$ but not $f_2$ (see Fig. \ref{fig3}).
The number of the elements for every kind is $2^{n^2-4}$. Thus, there are a total of $4\times2^{n^2-4}=2^{n^2-2}$ elements that are not commutative with $H_{ab}$, which transform $H_{ab}$ into $-\sigma_x^a\sigma_x^b-\sigma_y^a\sigma_y^b+\sigma_z^a\sigma_z^b$. Since the other $2^{n^2-2}$ elements leave $H_{ab}$ unchanged, we obtain the averaged operator: $\prod_{\mathcal{B}^z}(H_{ab})=\sum_{i=1}^{2n^2}\sigma_z^a\sigma_z^b$. The remaining terms can be wiped out by the group $\mathcal{B}^x$ because half of its elements are commutative with $\prod_{\mathcal{B}^z}(H_{ab})$ while the others are not, indicating the same situation as above.

Despite the significant decoupling power, the required steps in the above procedure scale exponentially with the number of the squares. This may limit its practical application. However, if only $H_{S\!E}$ and nearest-neighbor two-qubit interactions $H_{a,b}=\sigma_{\alpha_a}^a\otimes\sigma_{\alpha_b}^b$ ($a$, $b$ are the qubit indexes and $\alpha_a,\alpha_b=x,y,z$) are relevant, a much simpler scheme can be developed.

To this end, we introduce two decoupling groups $\mathcal{T}^z$ and $\mathcal{T}^x$.
Explicitly, $\mathcal{T}^z=\{I, t_1, t_2, t_1t_2\}$ and  $\mathcal{T}^x=\{I, t_1^d, t_2^d, t_1^dt_2^d\}$, where $t_1$ ($t_1^d$) is a Pauli operator chain which comprises the $\sigma_z$ ($\sigma_x$) operators acting on all the qubits attached on the vertical edges in the original (dual) lattice, and $t_2$ ($t_2^d$) comprises the $\sigma_z$ ($\sigma_x$) operators acting on all the qubits attached on the horizontal edges in the original (dual) lattice.
The topological meaning of the decoupling groups $\mathcal{T}^z$ and $\mathcal{T}^x$ is clear.
The four elements in group $\mathcal{T}^z$ ($\mathcal{T}^x$) are related to the topologically trivial circles, the circles surrounding the ``handle'', the circles encircling the genus, and the circles surrounding both the ``handle'' and the genus in the original (dual) lattice, respectively.

\begin{figure}[!tb]
  \centering
  \includegraphics[width=0.45\textwidth]{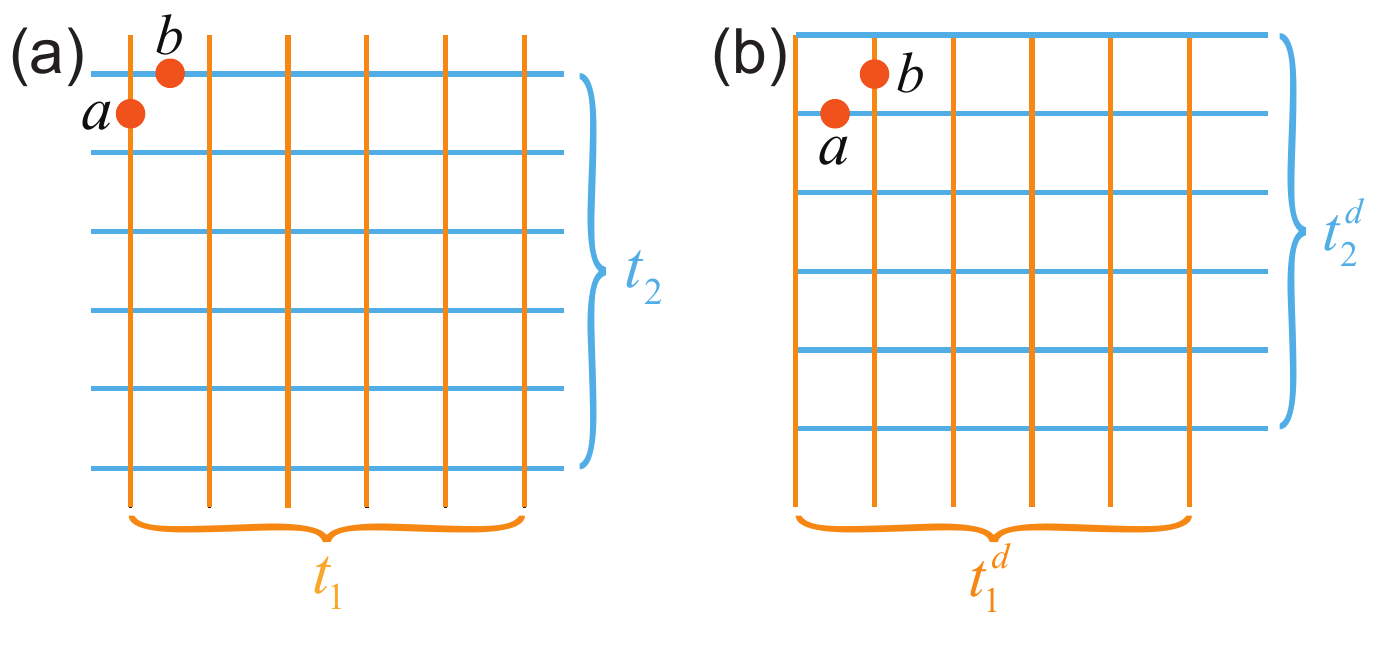}
  \caption{(a) Schematic diagram for the group element $t_1$ and $t_2$ in the original lattice. (b) Schematic diagram for the corresponding group element $t_1^d$ and $t_2^d$ in the dual lattice.}\label{fig4}
\end{figure}
Based on $\mathcal{T}^z$, we can design a decoupling procedure $D^z=[t_1t_2,\tau,t_1,\tau,t_2t_1,\tau,t_1,\tau,I]$, in which $t_i\cdot t_i=I$ ($i=1,2$) is used.
The corresponding dynamically averaged operator for $H_{S\!E}$ can be written as
\begin{equation}\label{}
   H_{{D}^z}=\prod_{\mathcal{T}^z}(H_{S\!E})=\sum_{i=1}^{2n^2}\sigma_z^i\otimes E_z^i,
\end{equation}
because each qubit on the lattice are associated with two elements in $\mathcal{T}^z$ which turn $\sigma_x^a\otimes E_x^a+\sigma_y^a\otimes E_y^a+\sigma_z^a\otimes E_z^a$ into $-\sigma_x^a\otimes E_x^a-\sigma_y^a\otimes E_y^a+\sigma_z^a\otimes E_z^a$ while the other two elements leave it unchanged.

Further, based on $\mathcal{T}^x$, a decoupling procedure, ${D}^{xz}=[t_1^dt_2^d,{D}^z,t_1^d,{D}^z,t_2^dt_1^d,{D}^z,t_1^d,{D}^z,I]$
can be established.
With ${D}^{xz}$, $H_{S\!E}$ can be removed from the dynamics of the qubits since
\begin{equation}\label{}
   H_{{D}^{xz}}=\prod_{{\mathcal{T}}^x}(H_{{D}^z})=0,
\end{equation}
for the same reason as $H_{{D}^z}$.

Besides $H_{S\!E}$, the nearest-neighbor interaction $H_{a,b}$ between two qubits can be decoupled too.
The key point is that, as shown in Fig.~\ref{fig4}, qubit $a$ is related to $t_1$ ($t_2^d$) and qubit $b$ is associated with $t_2$ ($t_1^d$) in the original (dual) lattice. It is easy to check that, just like $H_{S\!E}$, $H_{a,b}$ commutes with two elements but anti-commutes with the other two elements in $\mathcal{T}^z$ ($\mathcal{T}^x$).
Therefore, we can obtain
\begin{equation}\label{}
  \prod_{\mathcal{T}^x}(\prod_{\mathcal{T}^z}(H_{a,b}))=0,
\end{equation}
for any pair of nearest-neighbor $a$ and $b$.

So far, our decoupling procedures are designed based on the original and dual lattices embedded in a torus (with genus $G=1$).
A closed surface with a higher genus $G=k$,  can be  obtained by gluing $k$ tori together. It follows that the corresponding $\mathcal{B}^z$ and $\mathcal{B}^{x}$ are $k2^{n^2-1}$ ordered groups while $T^z$ and $T^x$  both have $4k$ elements.
Therefore, the decoupling groups on such a surface can be defined and similar decoupling procedures can be developed.

\begin{figure}
  \centering
  \includegraphics[width=0.4\textwidth]{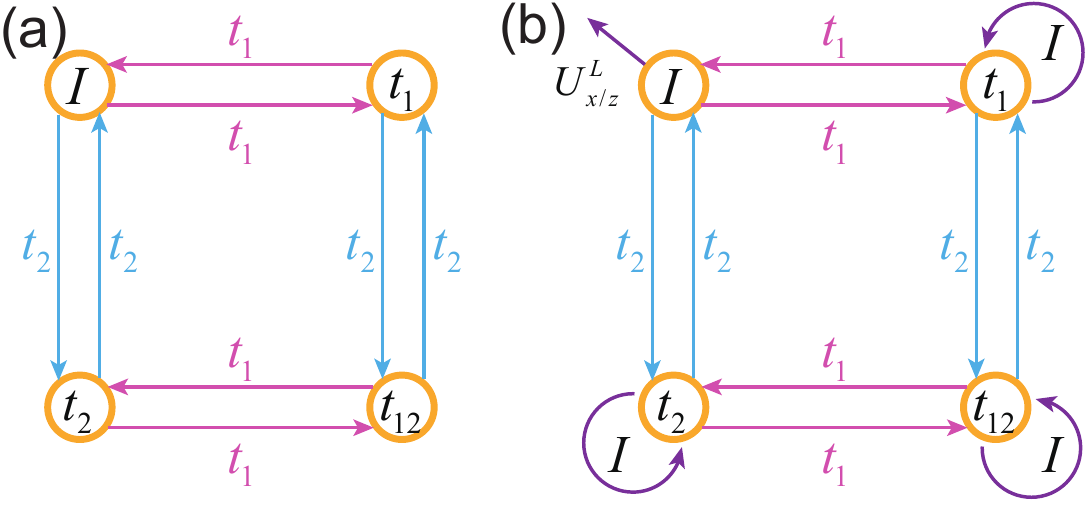}
  \caption{(a) Cayley graph and Eulerian cycles for $\mathcal{T}^z$. One possible Eulerian cycle is $I\rightarrow t_1\rightarrow t_{12}\rightarrow t_2\rightarrow I\rightarrow t_2\rightarrow t_{12}
\rightarrow t_1\rightarrow I$. (b) A modified Eulerian cycle and its Cayley graph to implement logical Pauli-$x$ and Pauli-$z$ operators. Three additional $I$ operators along with a logical operator generated with the same Hamiltonian are added to the original Eulerian cycles.}\label{fig5}
\end{figure}

\section{Realizing $D^z$ with bounded-strength controls}
In the above section, the decoupling procedure $D^z$ is developed by assuming
the decoupling operators $t_1$ and $t_2$ can be achieved instantaneously. This
requires the capability of applying arbitrarily strong control Hamiltonians,
which is not practical in experiments. Below, we show that the same effective
Hamiltonian $H_{D^z}$ can be realized with bounded-strength controls.

Our method is to use Eulerian cycles on the Cayley graph \cite{viola03} associated with group $\mathcal{T}^z$ (see Fig.~\ref{fig5}).
Since $\mathcal{T}^z$ consists of two generators $t_1$ and $t_2$, the corresponding Eulerian cycles comprise $8$ directed edges.
One of the possible control paths can be written as $I\xrightarrow{t_1}t_1\xrightarrow{t_2}t_{12}\xrightarrow{t_1}t_2\xrightarrow{t_2}I\xrightarrow{t_2}t_2\xrightarrow{t_1}t_{12}
\xrightarrow{t_2}t_1\xrightarrow{t_1}I$ [see Fig.~\ref{fig5}(a)], where the time interval between two adjacent decoupling operators is $\tau$.
Note that each decoupling operator in the path is related with its former one with a generator ($t_1$ or $t_2$), implying that we only have to repeatedly applying the generators in the time interval $\tau$ to the qubits according to the control path.
It follows that the needed control Hamiltonian for $t_1$ or $t_2$ can be
written as $H_{i}^z=\frac{\pi}{2\tau}\sum_{k\in S_i}\sigma_z^k$,
where $S_i$ being the set of qubits related with $t_i$.
It is clear that when $\tau$ is nonzero, $H_{1}^z$ and $H_2^z$ are
bounded-strength controls.
The associated dynamically averaged Hamiltonian $H^E_{D^z}$ is identical with $H_{D^z}$. It follows that if $H_{SE}$ consists of only $\sigma_{x/y}\otimes B_{x/y}$ terms, it can be fully decoupled from the total dynamics.

Given the same assumption, Eulerian cycles based on a modified Cayley graph allow us to build logical gates which are protected by the decoupling procedure [see Fig.~\ref{fig5}(b)].
The key idea is that when a particular logical gate $U^L$ can be constructed with a Hamiltonian $H^L(t)$, we can also obtain an identity with use of $H^L$ and with the same error as $U^L$ \cite{kho09,kho09pra}.
Thus, by adding an identity to each vertex in the Cayley graph (except the
one for $U^L$),
the error caused by $H_{S\!E}$ is averaged by group $\mathcal{T}^z$, leaving a net operator $U^L$ acting on the qubits.
Given a square lattice, there are two logical qubits that can be encoded. The corresponding logical $\sigma_z$ and logical $\sigma_x$ operators can be constructed with the modified Eulerian cycles [shown in Fig.~\ref{fig5}(b)].

We generalize the above idea to realize the procedure $D^{xz}$ whose Cayley
graph has 16 vertices with an Eulerian cycle consisting of 64 edges (see
Supplemental Materials). Higher order errors can also be eliminated by combining
this idea with concatenated DD \cite{kh05,kh10,west10}, so that the procedure can be implemented for long time decoupling.

\begin{figure}[!tb]
  \centering
  \includegraphics[width=0.45\textwidth]{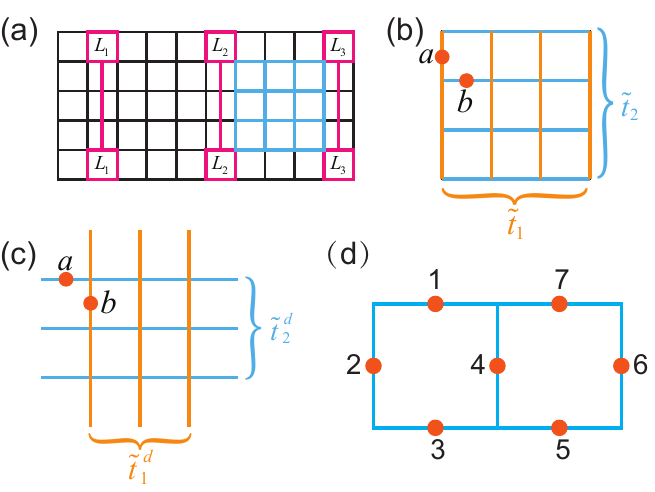}
  \caption{(a) A finite planar square lattice to implement surface codes. Each logical qubit is characterised by a pair of pink squares. When a logical operation on $L_1$ is performing, the physical qubits on the small lattice colored blue could be ``idle'' qubits. (b) The ``idle'' lattice shown in (a). The red solid circles on the edges are qubits $a$ and $b$, respectively. (c) The dual lattice of the $3\times3$ lattice in (b). (d) The $1\times2$ lattice considered in the numerical simulations.}\label{fig6}
\end{figure}
\section{Application to surface codes}
In the above discussion, we focus on the case where  the qubits are arranged on a periodic square lattice. Our method can be directly generalized to the case where the qubits are placed on a finite planar square lattice (i.e., a planar lattice with boundaries) shown in Fig.~\ref{fig6}(a). This is exactly the same lattice used to implement surface codes \cite{fowler12pra}.

In doing quantum computation with surface codes, the qubits belonging to some logical qubits (e.g., $L_1$) are manipulated  while the qubits in the rest area of the lattice, e.g., the qubits labeled blue in Fig.~\ref{fig6}(a), have to wait until a certain logical gate is performed.
This leaves the ``idle'' qubits interacting with the environment for a considerable long time, which can cause severe decoherence that may not be correctable \cite{stephens14}.
Besides, unwanted interactions between qubits can spread errors from one qubit to another, reducing the fault-tolerance of surface codes.
Therefore, a method that can suppress the environmental noises on the ``idle'' qubits, remove the unwanted interactions from the qubits' dynamics is strongly desired.

This task can be fulfilled by introducing two similar decoupling groups $\tilde{\mathcal{T}}^{z}$ and $\tilde{\mathcal{T}}^{x}$.
Here, $\tilde{\mathcal{T}}^{z}=\{I, \tilde{t}_1, \tilde{t}_2, \tilde{t}_1\tilde{t}_2\}$ and  $\tilde{\mathcal{T}}^{x}=\{I, \tilde{t}_1^{d}, \tilde{t}_2^{d}, \tilde{t}_1^{d}\tilde{t}_2^{d}\}$, where $I$ is the identity of the qubits,
$\tilde{t}_1$ ($\tilde{t}_1^{d}$) consists of the $\sigma_z$ ($\sigma_x$) operators acting on all the qubits attached on the vertical edges in the original (dual) lattice, and $\tilde{t}_2$ ($\tilde{t}_2^{d}$) comprises the $\sigma_z$ ($\sigma_x$) operators acting on the qubits attached on the horizontal edges in the original (dual) lattice.

Similarly, we can define two procedures $\tilde{D}^{z}=[\tilde{t}_1\tilde{t}_2,\tau,\tilde{t}_1,\tau,\tilde{t}_2\tilde{t}_1,\tau,\tilde{t}_1,\tau,I]$ and $\tilde{D}^{xz}=[\tilde{t}_1^{d}\tilde{t}_2^{d},\tilde{D}^{z},\tilde{t}_1^{d},\tilde{D}^{z},\tilde{t}_2^{d}\tilde{t}_1^{d},
\tilde{D}^{z},\tilde{t}_1^{d},\tilde{D}^{z},I]$ based on $\tilde{\mathcal{T}}^{z}$ and $\tilde{\mathcal{T}}^{x}$.
Based on $\tilde{\mathcal{T}}^{z}$, we develop a decoupling procedure
$\tilde{D}^{z}=[\tilde{t}_1\tilde{t}_2,\tau,\tilde{t}_1,\tau,\tilde{t}_2\tilde{t}_1,
\tau,\tilde{t}_1,\tau,I]$.
For a qubit on the lattice [e.g., qubit $a$ in Fig.~\ref{fig6}(b)], it is contained only in $\tilde{t}_1$ and $\tilde{t}_1\tilde{t}_2$. It follows that the interaction $H_{S\!E}^a$ between qubit $a$ and its environment is modified to $\prod_{\tilde{\mathcal{T}}^{z}}(H_{S\!E}^a)=\sigma_z^a\otimes E_z^a$. Thus
\begin{equation}\label{}
  H_{\tilde{D}^{z}}=\prod_{\tilde{\mathcal{T}}^{z}}(H_{S\!E})=\sum_{i=1}^{2(n^2+n)}\sigma_z^i\otimes E_z^i.
\end{equation}
After another decoupling procedure $\tilde{D}^{xz}=[\tilde{t}_1^{d}\tilde{t}_2^{d},\tilde{D}^{z},\tilde{t}_1^{d},\tilde{D}^{z},\tilde{t}_2^{d}\tilde{t}_1^{d},
\tilde{D}^{z},\tilde{t}_1^{d},\tilde{D}^{z},I]$ based on $\tilde{\mathcal{T}}^{x}$, $H_{S\!E}$ can be completely decoupled since each remaining $\sigma_z^i\otimes E_z^i$ term is anti-commutative with two elements (one is $\tilde{t}_1^d$ or $\tilde{t}_2^d$, the other is $\tilde{t}_1^d\tilde{t}_2^d$) in $\tilde{\mathcal{T}}^{x}$.

From Fig.~\ref{fig6}(b) and (c) we observe that a pair of nearest-neighbor qubits $a$ and $b$ is always associated with $\tilde{t}_1$ ($\tilde{t}_2^d$) and $\tilde{t}_2$ ($\tilde{t}_1^d$) in the original (dual) lattice, respectively.
This is the same case as we meet in the periodic lattice. Thus, the point there is also valid:
$H_{a,b}$ commutes with two elements but anti-commutes with the other two elements in $\tilde{\mathcal{T}}^z$ ($\tilde{\mathcal{T}}^x$), implying the same result $ \prod_{\tilde{\mathcal{T}}^x}(\prod_{\tilde{\mathcal{T}}^z}(H_{a,b}))=0$.

\section{Numerical simulations}
\begin{figure}[!tb]
  \centering
  \includegraphics[width=0.45\textwidth]{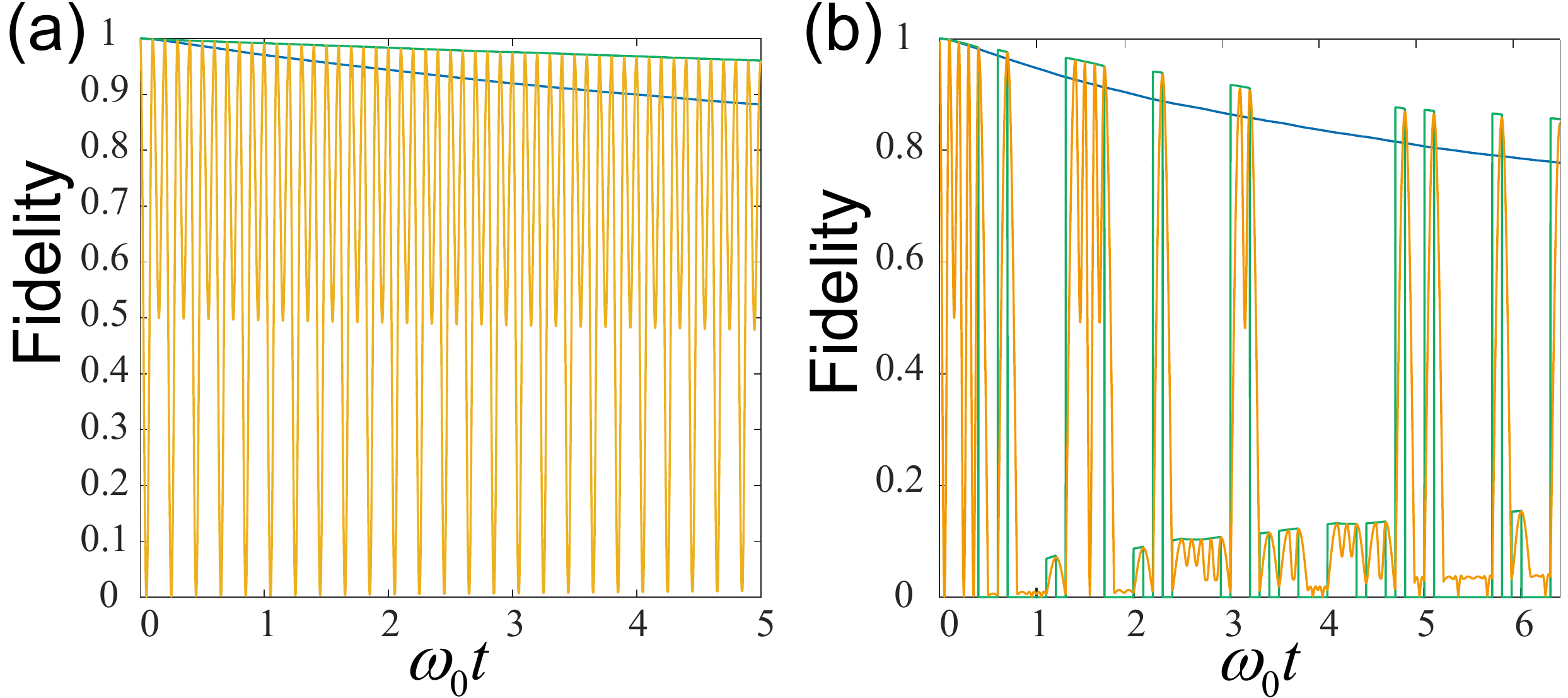}
  \caption{Fidelities between the initial state $\ket{\psi^L}$ and the corresponding $\rho(t)$ obtained from the reduced dynamics related to $D^z$ or $D^{xz}$ versus evolution time. For each procedure, we consider three cases: (i) with no control Hamiltonian (blue curves), (ii) with bounded-strength control Hamiltonians (yellow curves), and (iii) with arbitrary-strong control Hamiltonians (green curves). (a) Fidelities for $D^z$. (b) Fidelities for $D^{xz}$.}\label{fig7}
\end{figure}
To show the validity of our scheme, we perform numerical simulations to
calculate the fidelities between an initial logical state and its corresponding
one obtained from the reduced dynamics (with or without the dynamical
decoupling). Here we consider a $1\times2$ lattice, which consists of $7$
qubits [see Fig.~\ref{fig6}(d)].  The environment interacting with the qubits is described by a set of
boson models (i.e., $H_E=\sum_{l}\omega_la_l^\dag a_l$), through the
interaction Hamiltonian $H_{S\!E}=\sum_{k,l}\sigma_\alpha^k(g_l^ka_l+g_l^{k*}
a_l^\dag)$, where $a_l$ ($a_l^\dag$) is the annihilation (creation) operator
for the boson model of frequency $\omega_l$, and $|g_l^k|$ correspond to the
strength of the interaction, ranging from $0.01\omega_0$ to $0.03\omega_0$.
The Hamiltonian for the total system takes the form
\begin{equation}\label{}
  H=\sum_{(a,b)\in\mathcal{I}}\omega_{a,b}H_{a,b}+H_E+\sum_{k=1}^{7}\sum_{l}\sigma_{\alpha}^k(g_l^ka_l+g_l^{k*}a_l^{\dag}),
\end{equation}
where $\mathcal{I}$ is the set of nearest-neighbor qubit pairs [such as ($1,2$),
($2,3$), ($3,4$), and ($4,1$)], $\omega_{a,b}$ is the strength of $H_{a,b}$.

Based on the stabilizers defined on the lattice, the logical state can be chosen as
\begin{equation}\label{}
  \ket{\psi^L}=\frac{1}{2}(\ket{0000000}+\ket{1111000}+\ket{0001111}+\ket{1110111}).
\end{equation}
To show the decoupling procedure clearly, we first define the following decoupling operators involved in groups $\tilde{\mathcal{T}}^z$ and $\tilde{\mathcal{T}}^x$:
\begin{align}
  &\tilde{t}_1=\sigma_z^2\sigma_z^4\sigma_z^6, \tilde{t}_2=\sigma_z^1\sigma_z^3\sigma_z^5\sigma_z^7, \tilde{t}_1\tilde{t}_2=\sigma_z^1\sigma_z^2\sigma_z^3\sigma_z^4\sigma_z^5\sigma_z^6\sigma_z^7; \nonumber \\
  &\tilde{t}_1^d=\sigma_x^1\sigma_x^3\sigma_x^5\sigma_x^7, \tilde{t}_2^d=\sigma_z^2\sigma_z^4\sigma_z^6, \tilde{t}_{1}^d\tilde{t}_2^d=\sigma_x^1\sigma_x^2\sigma_x^3\sigma_x^4\sigma_x^5\sigma_x^6\sigma_x^7.
\end{align}

When $H_{SE}$ consists of only $\sigma_x$ and $\sigma_y$ terms, it can be
removed from the dynamics of the qubits with decoupling group $\mathcal{T}^z$.
In the first simulation, we choose that qubits $1$ to $4$ couple to the
environment with $\sigma_x$ and qubits $5$ to $7$ couple to the environment
with $\sigma_y$,
the time interval between two pulses to be $\tau=0.1\omega_0t$ and all
$\omega_{a,b}=0$.
We consider three cases: (i) free evolution without control, (ii) realizing
$D^z$ with arbitrarily strong controls, (iii) realizing $D^z$ with
bounded-strength controls. The results have been shown in  Fig.~\ref{fig7}(a). One may notice that the fidelity for case (iii) fluctuates with
time. This is because the initial state $\ket{\psi^L}$ is an eigenstate of the
decoupling operators, but not an eigenstate of the control Hamiltonians $H_1^z$
and $H_2^z$.

When $H_{SE}$ comprises $\sigma_x$, $\sigma_y$, and $\sigma_z$ terms at the same time, we need to use $D^{xz}$ as the decoupling procedure. In this case, the qubits interact with the environment through Pauli operators randomly.
In this simulation, we also set $\tau=0.1\omega_0t$.
Under this condition, $D^{xz}$ can be realized with bounded-strength controls based on Eulerian cycles (for details, see Supplemental Materials).
Here, we also set all $\omega_{a,b}=0$ and consider three cases: (i) free
evolution without control, (ii) realizing $D^{xz}$ with arbitrarily strong
controls, (iii) realizing $D^{xz}$ with bounded-strength controls. The results
have been shown in  Fig.~\ref{fig7}(b).

\begin{figure}
  \centering
  \includegraphics[width=0.4\textwidth]{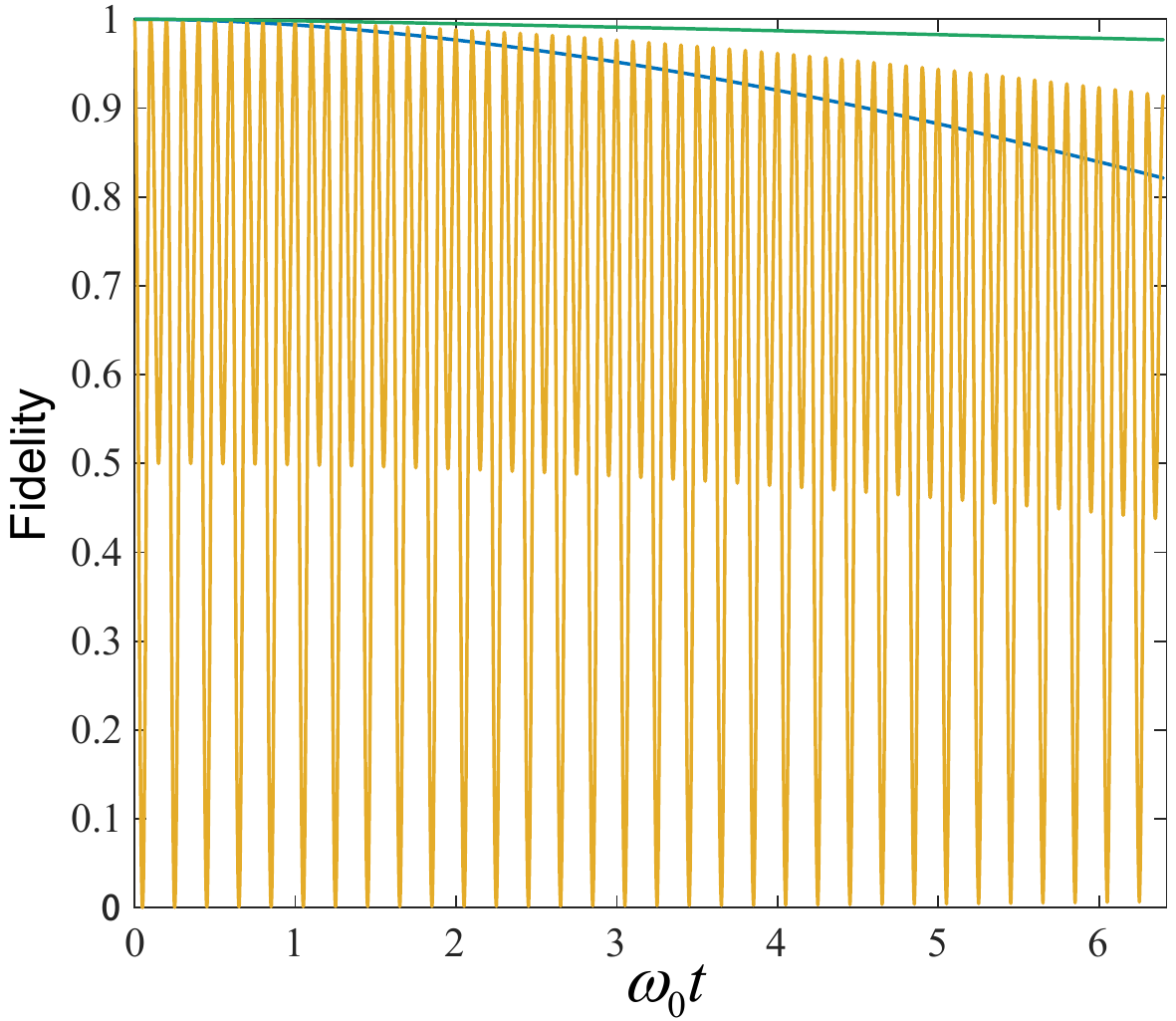}
  \caption{Fidelities between the initial state $\ket{\psi^L}$ and the
    corresponding $\rho(t)$ obtained from the reduced dynamics related to $D^z$
    with $\omega_{a,b}=0.03\omega_0$ versus evolution time. We consider three
    cases: (i) with no control Hamiltonian (blue curve), (ii) with
  bounded-strength control Hamiltonians (yellow curve), and (iii) with
arbitrarily strong control Hamiltonians (green curve).}\label{fig8}
\end{figure}

We also consider a case where $\omega_{a,b}$ are set to be $0.03\omega_0$ (i.e., the interactions between nearest-neighbor qubits are open) and $H_{SE}$ consists of only $\sigma_x$ and $\sigma_y$ terms. By using $D^z$ and its related Eulerian cycles (setting $\tau=0.1\omega_0t$), we consider three cases: (i) free evolution without control, (ii) realizing $D^z$ with arbitrarily strong controls, (iii) realizing $D^z$ with bounded-strength controls. The results are shown in Fig.~\ref{fig8}.

The reduced dynamics of the qubits can be obtained by using the stochastic
Liouville equation methods~\cite{sto97,sto02,koch08,yan16}. We choose a Lorenz
type spectrum $J(\omega)=\gamma/(\gamma^2+\omega^2)$ without the Matsubara
terms, where $\gamma$ is the inverse of the bath correlation time. The
corresponding correlation function is assumed to be of the exponential form
$\alpha(t-s)=(\Gamma\gamma/2)\exp(-\gamma|t-s|)$ with $\Gamma$ being the
coupling strength to the environment and $\gamma=1$.

The state fidelity is defined as $F(t)=\sqrt{\bra{\psi^L}\rho(t)\ket
{\psi^L}}$, where $\rho(t)$ is obtained from the reduced dynamics.
In Fig.~\ref{fig7} (a) and (b), fidelities for the decoupling procedure $D^z$ and $D^{xz}$ are demonstrated.
In order to show the instantaneous dynamics clearly, we choose a relatively long
inner-pulse interval $\tau=0.1\omega_0 t$. Despite this, the numerical results
show that our scheme improves the free fidelities considerably.
For example, Fig.~\ref{fig7}(a) shows that, compared with the final fidelity for the free evolution ($88.2\%$), the final fidelities for the ideal control and the Eulerian cycle increase by $7.8\%$ and $7.6\%$, respectively.
With a shorter $\tau$, the fidelities can be improved further.  Another significant advantage demonstrated in
Fig.~\ref{fig7} is that our scheme does not require arbitrary strong pulses:
for $D^z$, the final fidelity for the bounded-strength controls is almost
identical to the arbitrarily strong ones; for $D^{xz}$, the difference between
the final fidelity for the bounded-strength controls and that for the ideal
ones is less than $1\%$. This indicates that our scheme can be
implemented with practical experimental technology.

\section{Conclusions}
In summary, we have used topological concepts to develop DD procedures to protect the qubits arranged on square lattices.
In each procedure,  the decoupling groups are formed from the original lattice and its dual.
Owing to the topological nature of the decoupling groups, quantum information stored in the homology degrees of freedom can be preserved.
We further show that the designed decoupling procedure can be implemented with realistic strength controls for a long time period.
As an example, we explicitly show how our scheme can be generalized to a practical surface code implementation where a planar lattice with boundaries is involved.
Our scheme opens a window to introduce DD approach to surface codes so that the errors caused by environments can be reduced, making the required threshold a more easily reachable target.

\begin{acknowledgments}
This work is supported by the National Basic Research Program of China under
Grant Nos. ~2017YFA0303700 and ~2015CB921001, National Natural Science
Foundation of China under Grant Nos.~61726801, ~11474168 and ~11474181, and in
part by the Beijing Advanced Innovative Center for Future Chip(ICFC). J.Z.
acknowledges support by the China Postdoctoral Science Foundation (Grant No.
2018M631437).
X.D.Y. acknowledges support by the DFG and the ERC (Consolidator Grant
683107/TempoQ).
\end{acknowledgments}

\end{document}